\documentstyle[12pt,epsf]{article}
\def\Sp{\mbox{Sp}\, }
\def\erfc{\mbox{erfc}}
\def\Re{\mbox{Re}\, }
\def\Im{\mbox{Im}\, }
\begin{document}
\baselineskip=17pt
{\large
\title{\bf Dynamical symmetry breaking in the Nambu-Jona-Lasinio model
with external gravitational and constant electric fields }
\author{E. Elizalde$^{1,2,}$\thanks{E-mail: eli@zeta.ecm.ub.es,
elizalde@io.ieec.fcr.es}, \  
Yu. I. Shil'nov$^{1,3,}$\thanks{Present e-mail: visit2@ieec.fcr.es} \ and 
V. V. Chitov$^{3,}$\thanks{E-mail: chitov@io.itl.net.ua}
\\[1.5ex]
$^1${\it Consejo Superior de Investigaciones Cient\'{\i}ficas,}\\
{\it IEEC, Edifici Nexus-204, Gran Capit\`a 2-4, 08034, Barcelona, Spain}\\
[0.5ex]
$^2${\it Department  ECM, Faculty of Physics, University of Barcelona,}\\
{\it Diagonal 647, 08028, Barcelona, Spain}\\
[0.5ex]
$^3${\it Department of Theoretical Physics, Faculty of Physics,}\\
{\it Kharkov State University, Svobody Sq. 4, 310077, 
Kharkov, Ukraine}}

\date{ }
\maketitle
\begin{abstract}
An investigation  of the Nambu-Jona-Lasino model 
with external constant electric
and weak gravitational fields  is carried out in three- and
four-dimensional
spacetimes. The effective potential of the
composite bifermionic fields is calculated keeping terms linear
in the curvature,
 while the  electric field effect is treated exactly
by means of the proper-time formalism.
A rich dynamical symmetry breaking pattern, accompanied by
phase transitions which are ruled, independently,
 by both the curvature and the electric field strength
is found. Numerical simulations of the transitions are presented.

\end{abstract}

\pagebreak
}

\section{Introduction.} The Nambu-Jona-Lasinio (NJL) model
\cite{NJL} has long been considered one 
of the most suitable approximations providing a
correct description of low-energy strong interactions physics.
It has been seen that  dynamical symmetry breaking 
(DSB)\footnote{This phenomenon and its application in high energy 
physics have been described in \cite{DSB, Mir}.}  and  dynamical
fermions mass generation  take place in the models with
four-fermion interactions, those of NJL and Gross-Neveu 
\cite{GN}.

External conditions, such as a non-zero temperature, a finite
chemical potential or classical gauge fields, have
been  shown to enrich the phase structure of the NJL model essentially
(for  reviews  and  complete lists of references
see, for instance, \cite{RWP}-\cite{naf}).

The influence of external  magnetic and electric fields in these models
 has been studied for some time \cite{Sch}-\cite{naf}. It has 
been found out that a  non-zero  magnetic field always breaks  the chiral
symmetry,  while the presence of an electric field (EF)  tries to restore
it.

Investigations of the influence of  a classical gravitational field on the
DSB phenomenon in the NJL model have been carried out for some years.
It has been shown that curvature-induced phase transitions exist 
and that they must be
incorporated  into any realistic scenario of the early Universe
 \cite{HS}-\cite{ELOS} (for a general introduction to quantum
field theory in curved spacetime see \cite{BOS} and for a
recent review of the NJL model  in curved spacetime, \cite{REV}).
It turns out that, in spite of the relatively small value of the
 curvature-dependent
corrections at the low energy scale to be investigated within
the NJL model, these corrections appear to be inescapable, in the sense 
that they must be taken into account when one performs
the necessary "fine tuning" of the different cosmological parameters
\cite{Linde}.

On the other hand, an external electromagnetic field
plays an important  role in the early Universe, which both can contain
primodial magnetic fields \cite{TW, MF} and have  a very large 
electrical conductivity \cite{TW, CON}. Investigations of the 
influence of an external
magnetic field within the DSB  in curved spacetime have been
carried out recently  \cite{GOS}-\cite{IOS}. Therefore, it seems
natural to study the effect of a constant EF upon the DSB
phenomenon in a curved spacetime extending the results of 
 \cite{KL, K1, JpEF}, which have been done for
the case of flat  spacetime.

In this paper we investigate the behaviour of the
 NJL model in the presence of an external constant
EF treated nonperturbatively in the proper-time
formalism \cite{Sch}. The linear-curvature corrections to the effective
potential (EP)
of  composite bifermionic fields are  calculated and the phase
structure of the model is investigated for negative values of the
 coupling constant.
 Both the  four-  and threedimensional cases are discussed. It is well known
that in an
external EF particle creation takes place and that the
EP acquires
an imaginary part \cite{Sch}-\cite{K1}. That is why we study
the small EF limit when the particle creation 
velocity is negligible.       
The phase transitions accompanying the DSB process 
on the spacetime curvature, as well as the values of EF strength
 will be described numerically in detail.

\section{The fermion Green function in an external constant
electromagnetic
field with linear-curvature accuracy.}

Our starting point is the Nambu-Jona-Lasinio model in
curved space-time of arbitrary dimension, $d$,  as
 given by the following action
\cite{NJL} :
$$
S=\int d^d x \sqrt{-g} \left\{i \overline{\psi}\gamma^\mu (x)D_\mu \psi +
{\lambda \over 2N} \biggl[ (\overline{\psi}\psi)^2+
(\overline{\psi} i \gamma_5 \psi)^2 \biggr] \right\},
\eqno(1)
$$
where the covariant derivative $D_{\mu}$ includes the electromagnetic
potential $A_{\mu}$:
$$ 
D_{\mu}=\partial_{\mu} -i e A_\mu+{1 \over 2 }\omega^{ab}{}_{\!\mu}
\sigma_{ab}. 
\eqno(2)
$$
The local Dirac matrices $\gamma_\mu (x)$ are expressed through the usual
flat ones $\gamma_a$ and the tetrads $e^a_\mu$:
$$
\gamma^\mu (x)=\gamma^a e^\mu_a (x),
\eqno(3)
$$
$$
\sigma_{ab}={1\over 4 }[\gamma_a,\gamma_b].
\eqno(4)
$$
 The spin connection has the form
$$
\omega^{ab}{}_{\!\mu}=\frac{1}{2}e^{a
\nu}(\partial_{\mu}e^{b}_{\nu}-\partial_{\nu}
e^{b}_{\mu})+\frac{1}{4}e^{a
\nu}e^{b\rho}e_{c\mu}(\partial_{\rho}e^{c}_{\nu}
-\partial_{\nu} e^{a}_{\rho})
$$
$$
-\frac{1}{2}e^{b\nu}(\partial_{\mu}e^{a}_{\nu}-\partial_{\nu}
e^{a}_{\mu})-
\frac{1}{4}e^{b\nu}e^{a\rho}e_{c\mu}(\partial_{\rho}e^{c}_{\nu}-
\partial_{\nu}e^{c}_{\rho}),
\eqno(5)
$$
where $N$ is the number of bispinor fields $\psi_a$. The spinor
representation dimension is supposed to be four. Greek and Latin indices
correspond to the curved and flat tangent spacetimes, respectively.

Introducing the auxiliary fields:
$$
\sigma=-{\lambda \over N }(\overline{\psi} \psi ), \qquad
\pi=-{\lambda\over N}
\overline{\psi} i \gamma_5 \psi ,
\eqno(6)      
$$
we can rewrite the action (1) as:
$$
S=\int d^d x \sqrt{-g} \ \left\{ i\overline{\psi}\gamma^\mu D_\mu \psi -
{N \over 2\lambda
}(\sigma^2+\pi^2)-\overline{\psi}(\sigma+i\pi\gamma_5)\psi \right\}.
\eqno(7)
$$
Then, the effective action in the large-$N$ expansion is given by:
$$
{1 \over N } \Gamma_{eff}(\sigma,\pi)=
-\int d^d x \sqrt{-g}\, {\sigma^2+\pi^2 \over 2\lambda} -
i\ln \det \left\{i\gamma^\mu (x)D_\mu-(\sigma+i\gamma_5\pi) \right\}.
\eqno(8)
$$
Here we can put $\pi=0$, because the final expression will depend on
the combination $\sigma^2+ \pi^2$ only. This means  that we are actually
considering  the Gross-Neveu model \cite{GN}.

Defining the EP as
$V_{eff} = -\Gamma_{eff}/ N\, \int d^d x\sqrt{-g}$, for constant
configurations of $\sigma$ and $\pi$  we obtain:
$$
V_{eff}={\sigma^2 \over 2\lambda }+i \Sp \ln \langle x| [ i\gamma^\mu
(x)D_\mu -\sigma] |x \rangle
\eqno(9)
$$
By means of the usual Green function (GF), which obeys the equation
$$
(i \gamma^\mu D_\mu-\sigma)_x G(x,x',\sigma)=\delta(x-x'),
\eqno(10)   
$$
we obtain the following formula
$$
V_{eff}'(\sigma)={ \sigma \over \lambda }-i \Sp G(x,x,\sigma).
\eqno(11)
$$
To calculate the liner curvature corrections 
 the local momentum
expansion formalism is the most convenient one \cite{LME}. In the
special Riemannian normal coordinate framework we have:
$$
g_{\mu\nu}(x)=\eta_{\mu\nu}-{1\over 3 } R_{\mu\rho\sigma\nu}y^\rho
y^\sigma,
\eqno(12)
$$
$$
e^\mu_ a (x)=\delta^\mu_a+{1\over 6 } R^\mu{}_{\!\rho\sigma a}y^\rho
y^\sigma,
\eqno(13)
$$
$$
\omega^{ab}{}_{\!\mu}\sigma_{ab}={1\over 2 }R^{ab}{}_{\!\mu\lambda}
y^\lambda\sigma_{ab},
\eqno(14)
$$
$$
y=x-x'.
\eqno(15)
$$
The vector potential of the external electromagnetic field is chosen to be
of the form:
$$
A_\mu(x)=-{1 \over 2 } F_{\mu\nu} x^\nu ,
\eqno(16)
$$
where $F_{\mu\nu}$ is the constant matrix of the electromagnetic field
strength tensor. Substituting Eqs. (12)-(16) into Eq. (10), we arrive at
the following equation for the GF:
$$
\biggl[ i\gamma^a (\delta^\mu_a+{1\over 6}R^\mu{}_{\!\rho\sigma
a}y^\rho y^\sigma)(\partial_\mu+
{1 \over 4}R_{bc\mu\lambda}y^\lambda\sigma^{bc}-ieA_\mu)-
\sigma\biggr] G(x,x',\sigma)=\delta(x-x')
\eqno(17)
$$
Fulfilling the expansion on the different spacetime curvature monomials 
$$
G=G_0+G_1+. . . ,
\eqno(18)
$$
where  $G_0$ is the GF in  flat spacetime,
$G_1 \sim R$, and so on, we obtain the iterative sequence of
equations:
$$
\biggl[ i\not{\!\partial}+e\not{\!\!A(x)}-\sigma\biggr]
G_0(x,x')=\delta(x-x')
\eqno(19)
$$
$$
\biggl[ i\not{\!\partial}+ 
e\not{\!\!A}(x)-\sigma\biggr] G_1(x,x',\sigma)+
\biggl[{i\over 6}\gamma^a R^\mu{}_{\!\rho\sigma a}y^\rho y^\sigma 
(\partial_\mu-i eA_\mu (x))
$$
$$
+{i\over 4}\gamma^a R_{bca\lambda} y^\lambda \sigma^{bc} \biggr]
G_0(x,x',\sigma)=0.
\eqno(20)
$$
Here and below we can already forget about the difference between the
two kinds
of indices, Greek and Latin, because it only shows up beyond the linear
curvature approximation in Eq. (20).
We introduce the factor:
$$
\Phi(x,x')=\exp  \left[ie\int^x_{x'} A^\mu (x'') dx''_{\mu} \right],
\eqno(21)
$$
which is the solution of the equation
$$
(\partial_\mu-ieA_\mu)_x\Phi(x,x')=0
\eqno(22)
$$
and assume that, just as for flat spacetime, the GF has the form
\cite{Sch}:
$$
G(x,x',\sigma)=\Phi(x,x')\tilde{G}(x-x',\sigma).
\eqno(23)
$$
Then, the evident dependence on $A_{\mu}(x)$ disappears and we obtain the
equation determining the GF $\tilde{G_1}(x-x',\sigma)$: 
$$
(i\not{\!\partial}-\sigma)_x\tilde{G_1}(x-x',\sigma)=-{i \over 6
}\gamma^a
R^{\mu}{}_{\!\rho\sigma a} y^{\rho} y^\sigma \partial_\mu
\tilde{G_0}(x-x',\sigma)
$$
$$
-{i \over 4}\gamma^a \sigma^{bc} R_{bca\lambda}y^\lambda
\tilde{G_0}(x-x',\sigma).
\eqno(24)
$$

The GF for the case of an external electromagnetic field
in flat spacetime is supposed to be known.
Denoting by $ G_{00}(x-x', \sigma)$  the GF that satisfies the
equation:
$$
(i\not{\!\partial}-\sigma)G_{00}(x-x',\sigma)=\delta(x-x'),
\eqno(25)
$$
we get:
$$
\int dx'' G_{00}^{-1}(x-x'',\sigma)\tilde{G_1}(x''-x',\sigma)=
$$
$$
-{i \over 6 } \gamma^a R^\mu{}_{\!\rho\sigma
a}(x-x')^\rho(x-x')^\sigma
 \partial_\mu^x\tilde{G_0}(x-x',\sigma)
$$
$$
-\frac{i}{4}\gamma^a\sigma^{bc}R_{bca\lambda}(x-x')^\lambda
 \tilde{G_0}(x-x',\sigma).
\eqno(26)
$$
Finally,
$$
\tilde{G}_1(x-x',\sigma)=\int dx''G_{00}(x-x'',\sigma)
$$
$$
\times\biggl[ -{i \over 6}
\gamma^a R^\mu{}_{\!\rho\sigma a}(x''-x')^\rho
(x''-x')^\sigma\partial_\mu^{x''}
\tilde{G}_0(x''-x',\sigma)
$$
$$
-{i \over 4}\gamma^a
\sigma^{bc}R_{bca\lambda}(x''-x')^\lambda\biggr]
\tilde{G}_0(x''-x',\sigma).
\eqno(27)
$$
 
We actually need the coincidence limit $x \to x'$ in order to calculate
the EP (11). This provides us with the opportunity to
simplify (27), especially for constant curvature spacetime,
where
$$
R_{\mu\sigma\kappa\lambda}={R \over d(d-1)} \biggl(\eta_{\mu\kappa}
\eta_{\sigma\lambda}-\eta_{\mu\lambda}\eta_{\kappa\sigma} \biggr).
\eqno(28)
$$
Thus, our basic expression for the GF $\tilde{G_1}$ in a
space-time of arbitrary dimension $d$, is the following:
$$
\tilde{G_1}(0,\sigma)=-\frac{iR}{12d(d-1)}\int dz G_{00}(-z,
\sigma)
\biggl[ 2\!\not{\!z}z^\mu\partial_\mu \tilde{G_0}(z,\sigma)
$$
$$
-2z^2\gamma^\mu \partial_\mu \tilde{G_0}(z,\sigma)+3(d-1)\!\not{\!z}
\tilde{G_0}(z,\sigma)].
\eqno(29)
$$
This expression can be substituted into Eq. (11) directly, because
$\Phi(x, x)=1$.
  
\section{Effective potential and phase transitions in the NJL model
in curved spacetime.}
 
The GF for the case of flat spacetime is found to be in the
proper-time representation \cite{Sch, K1, GMS}:
$$
\tilde{G_0}(z,\sigma)=e^{-i {\pi d/4}}\int\limits_0^\infty
\frac{ds}{(4\pi s)^{d/2}}
e^{-is\sigma^2} \exp(-\frac{i}{4s}z_\mu C^{\mu\nu}z_\nu)
$$
$$
\times\biggl(\sigma+\frac{1}{2s}\gamma^\mu C_{\mu\nu}z^{\nu}-
\frac{e}{2}\gamma^\mu F_{\mu\nu}z^\nu\biggr)
\biggl[\tau \coth \tau-\frac{es}{2}\gamma^\mu\gamma^\nu
F_{\mu\nu}\biggr],
\eqno(30)
$$
$$
G_{00}(-z,\sigma)=e^{-i\pi d/4} \int\limits_0^\infty
{dt \over (4\pi t)^{d/2}} exp\biggl[-i(\sigma^2 t+{z^2 \over
4t})\biggr]
\biggl(\sigma +{1\over 2t}\gamma^{\mu}z_{\mu}\biggr),
\eqno(31)
$$
where:
$$
C_{\mu\nu}=\eta_{\mu\nu} -F_{\mu}{}^{\lambda} F_{\lambda\nu}
{1-(eEs)\coth(eEs) \over E^2}.
\eqno(32)
$$

Let us now consider the 3D case.
After a Wick rotation $is \rightarrow s, \,\, it\rightarrow t$,
we have the following expression:
$$
\Sp \tilde{G}_1(0,\sigma)={iR\sigma \over 72\pi^{3/2}} \int dtds
{e^{-(t+s)\sigma^2} \over (t+s)^{3/2}(1+\kappa\cot\tau)^2}
$$
$$
\times\biggl[-2\kappa(\kappa+\tau)+(9\tau+5\kappa)\cot\tau
+\kappa(\tau-3\kappa)\cot^2\tau\biggr],
\eqno(33)
$$
where $\tau= eEs, \kappa=eEt$. 
Performing the  integration  over $\sigma$, we get the EP:
$$
V_{eff}(\sigma)={\sigma^2\over 2\lambda}+ {1\over 4\pi^{3/2}}
\int\limits_{1/\Lambda^2}^\infty {ds\over s^{5/2}}
e^{-s\sigma^2}\tau \cot\tau
$$
$$
-{R\over 144\pi^{3/2}}\int\limits_{1/\Lambda^2}^\infty
\int\limits_{1/\Lambda^2}^\infty
{ds dt e^{-(t+s)\sigma^2} \over(t+s)^{5/2}(1+\kappa
\cot\tau)^2}
$$
$$
\times \biggl[-2\kappa(\kappa+\tau)+(9\tau+5\kappa)\cot\tau+
\kappa(\tau-3\kappa)\cot^2\tau\biggr].
\eqno(34)
$$
This expression coincides exactly, in the limit $E\to 0$,
 with previous results \cite{GOS}:
$$
V_{eff}^{(E=0)}(\sigma) =
\frac{\sigma^2}{2\lambda}+\frac{{\Lambda}^3}{6{\pi}^{3/2}}
\left\{ (1-2\frac{\sigma^2}{\Lambda^2})
\exp ( -{\sigma^2 \over\Lambda^2})+
2\sqrt{\pi}\frac{\sigma^3}{\Lambda^3}
\erfc (\frac{\sigma}{\Lambda})\right.
$$
$$
-\left. \frac{R}{4\Lambda^2}\left[ \exp ( -{\sigma^2 \over \Lambda^2}) -
\sqrt{\pi}\frac{\sigma}{\Lambda}
 \erfc (\frac{\sigma}{\Lambda})\right]\right\},
\eqno(35)
$$
where
$$
\erfc (x)=\frac{2}{\sqrt\pi}\int\limits_x^\infty e^{-t^2}dx.
\eqno(36)
$$

There are two ways of justifying the introduction of the $\Lambda$ parameter
in equations (34) and (35) for the EP.
The first one is the standard renormalization procedure, by means of
 the  UV cutoff
method. Then, in the limit $\Lambda\to\infty$, after the renormalization of
the coupling constant
$$
\frac{1}{\lambda_R}=\frac{1}{\lambda}-\frac{\Lambda}{\pi^2},
\eqno(37)
$$
we have the well known   expression for the 
 renormalized EP in  flat
spacetime \cite{K1,K2} with linear curvature corrections \cite{GOS}:
$$
V_{eff, R}^{E=0}(\sigma)=
\frac{\sigma^2}{2\lambda_R}+\frac{\sigma^3}{3\pi}+\frac{R\sigma}{24\pi}.
\eqno(38)
$$ 
Three-dimensional four-fermions models have been shown to be  renormalizable in 
the leading large-$N$ order \cite{RWP}. Furthemore, external electromagnetic and
gravitational fields do not interfere with  the renormalization procedure,
because the only divergence that
appears in the expression for the EP
(34) has already been removed by the substitution (37). This is due to the
obvious fact that the local feature of renormalizability cannot be spoiled
by the weak curvature of global spacetime or by an external EF.
Formally this can be proved by looking at the calculations of the leading
terms of the integrand in
the limit $s\to 0, t\to 0$, which are the only  essential ones in order
 to determine
the  UV-divergences of the EP (34). Thus, the renormalization has been
performed by the formula (37) completely.

However, in general,  four-fermion models may be considered  as
low energy effective theories being derived from a more
complete version of a quantum field theory (QCD, for example).
 In this case
the parameter $\Lambda$ can be treated as a natural characteristic
scale limiting the range where our low-energy approximation is valid,
and then our  model will
describe some phenomenological effects of elementary particle physics.
Bearing  these reasons in mind, we can otherwise maintain
 $\Lambda$ finite and
investigate the DSB phenomenon in our model with this fixed  cut-off.

The integrand contains the function
$\cot(eEs)$, that periodically goes to $\infty$.
 It  developes an infinite set of poles on the integration path
and obligues us to take into account the contribution of the
corresponding residua, given by the imaginary
part of the EP. The presence of imaginary terms  in 
the EP means that  particle
creation takes places and that our vacuum is actually unstable. 
The solution to
this problem lies outside the limits of our present investigation.
The simplest possibility
seems to be to consider a comparatively small EF
strength, with specific values
that would provide an exponentially depressed particle creation rate.

To start with we shall consider the case of a 3D flat spacetime:
$$
V_{eff}^{flat}(\sigma)={\sigma^2\over 2\lambda} +
{1 \over 4\pi^{3/2}}\int\limits_{1/\Lambda^2}^\infty
{ds\over s^{5/2}} e^{-s \sigma^2} \tau \cot\tau,
\eqno(39)
$$
where the integrand function has its poles at the points
$$
s_n={\pi n\over eE}, \ \  n=1,2,...
\eqno(40)
$$
The standard residue technique yields:
$$
V_{eff}^{flat}(\sigma)={\sigma^2\over 2\lambda}+
{1 \over 4\pi^{3/2}} P.v.\int\limits_{1/\Lambda^2}^\infty
{ds\over s^{5/2}} e^{-s \sigma^2} \tau \cot\tau
$$
$$
-i{(eE)^{3/2} \over 4\pi^2} \sum\limits_{n=1}^\infty
n^{ -3/2}\exp( -{\pi n\over eE}\sigma^2),
\eqno(41)
$$
where P.v. means ``principal value.''
Using the analytical continuation of the Hurwitz $\zeta$-function,
we find the following expression for the renormalized EP:
$$
V_{eff, R}^{flat}(\sigma)={\sigma^2\over 2\lambda_R}-
{(2ieE)^{3/2} \over 4\pi}
\biggl[ 2 \zeta (-{1\over2},{\sigma^2 \over 2ieE})-
\biggl({\sigma^2\over 2ieE}\biggr)^{1/2} \biggr].
\eqno(42)
$$

A numerical analysis of
$\Re V_{eff, R}^{flat}(\sigma)$ 
for negative coupling constant gives the typical behaviour of a
first--order phase transition, as shown
in figure 1. The critical values are defined as usual:
$E_{c1}$ corresponds to the strength of EF  for which
a local nonzero minimum appears,  $E_{c}$, when the real part of EP
is equal at zero and at the local minimum, and  $E_{c2}$, when the zero
extremum becomes a maximum.
For all figures,
an arbitrary dimensional parameter, $\mu$, defining a typical scale in
the model, is introduced in order to perform the plots in terms of
 dimensionless variables. 
$\Lambda$ obviously does not appear anywhere because after renormalization
(37) it must be set such that $\Lambda\to\infty$.
   
The curvature-dependent part of the effective potential does not 
contain  new UV-divergences
but, unfortunately, $\Lambda$ must be kept finite here, 
 because the integrals cannot
be calculated in the general case. Therefore, an analytical continuation 
cannot be perfomed. However, as  mentioned above, the limit of weak
electric field is the most suitable one here in order to investigate DSB
correctly. The most natural way is to keep the flat part of EP  in the
same nonpertubative form (42) meanwhile in the curvature-dependent terms
which are small already by themselves only the weak EF limit is taken into 
account. Thus, neglecting
the next order exponentially depressed terms in curvature corrections,
the renormalized EP turns out to be the following:
$$
V_{eff, R}^{(3D)}(\sigma)={\sigma^2\over 2\lambda_R}-
{(2ieE)^{3/2} \over 4\pi}
\biggl[ 2 \zeta (-{1\over2},{\sigma^2 \over 2ieE})-
\biggl({\sigma^2\over 2ieE}\biggr)^{1/2} \biggr]
$$
$$
+{R\sigma \over 24\pi}+{iR(eE)^{1/6} \over 2\pi^2 3^{7/3}}
\exp(-\pi{\sigma^2 \over eE})\Gamma(\frac{2}{3})\sigma^{2/3}.
\eqno(43)
$$
The shape of $\Re V_{eff, R}^{(3D)}(\sigma)$ is shown in figure 2, for the
case of a fixed curvature and coupling constant, and in figure 3, for the
case of fixed EF
strength and coupling constant. The shapes are absolutely typical
of first-order phase transition pictures. 

For a finite cut-off scale $\Lambda$, the effective potential is given by:
$$
\Re V_{eff}^{(3D)}(\sigma)= {\sigma^2 \over 2 \lambda}+
{1 \over 4\pi^{3/2}} P.v.\int\limits_{1/\Lambda^2}^\infty
{ds\over s^{5/2}} e^{-s \sigma^2} \tau \cot\tau
$$
$$
-{R\over 144\pi^{3/2}} P.v. \int\limits_{1/\Lambda^2}^\infty
\int\limits_{1/\Lambda^2}^\infty
{{ds dt e^{-(t+s)\sigma^2}} \over {(t+s)^{5/2}(1+\kappa \cot\tau)^2}}
$$
$$
\times \biggl[-2\kappa(\kappa+\tau)+(9\tau+5\kappa)\cot\tau+
\kappa(\tau-3\kappa)\cot^2\tau\biggr],
\eqno(44)
$$
$$
\Im V_{eff}^{(3D)}= 
-{(eE)^{3/2} \over 4\pi^2} \sum\limits_{n=1}^\infty
n^{ -3/2}\exp ( -{\pi n\over eE}\sigma^2)
$$
$$
+{R\over 144}({eE\over \pi})^{1/2} \sum\limits_{n=1}^\infty
\exp( -{\pi n\over eE}\sigma^2) 
\int\limits_{eE/\Lambda^2}^\infty dy
{\exp \left( \frac{\sigma^2}{eE} 
(\varphi-y)\right) \over {(1+y^2)^2
(y+\pi n -\varphi)^{5/2}}}
$$
$$
\times \biggl[ 2y(y^2+4)(\pi n -\varphi +y)
{\sigma^2\over eE}+
3(\pi n-\varphi +y)(3-y^2) +y(8-y^2)\biggl],
\eqno(45)
$$
where
$$
\varphi=\arccos {1\over \sqrt{1+y^2}}.
\eqno(46)
$$

In contrast to the case of $D=3$,
the 4D four-fermion models are not renormalizable.
Thus, we have to keep $\Lambda$ finite,
and do just the same type of calculations as for the previous 3D situation:
$$
\Sp \tilde{G}_1(0,\sigma)={iR\sigma \over 96\pi^2} \int dtds
{e^{-(t+s)\sigma^2} \over (t+s)^2 (1+\kappa\cot\tau)^2}
$$
$$
\times\biggl[-\kappa(\kappa+\tau)+2(\kappa+3\tau)\cot\tau +
2\kappa(\tau-\kappa)cot^2\tau\biggr].
\eqno(47)
$$
We  arrive at  the EP with a finite cut-off scale:
$$
\Re V_{eff}^{(4D)}= {\sigma^2 \over 2 \lambda}+
{1 \over 8\pi^2} P.v.\int\limits_{1/\Lambda^2}^\infty
{ds\over s^3} e^{-s \sigma^2} \tau \cot\tau
$$
$$
-{R\over 192\pi^2} P.v. \int\limits_{1/\Lambda^2}^\infty
\int\limits_{1/\Lambda^2}^\infty
{{ds dt e^{-(t+s)\sigma^2}} \over {(t+s)^3 (1+\kappa \cot\tau)^2}}
$$
$$
\times \biggl[-\kappa(\kappa+\tau)+2(\kappa+3\tau)\cot\tau+
2\kappa(\tau-\kappa)\cot^2\tau\biggr],
\eqno(48)
$$
$$
\Im V_{eff}^{(4D)}=-{(eE)^2\over 8\pi^3} \sum\limits_{n=1}^\infty
n^{-2}\exp( -{\pi n\over eE}\sigma^2)
$$
$$
+{R (eE) \over 192 \pi} \sum\limits_{n=1}^\infty
\exp(-{\pi n\over eE}\sigma^2)
\int\limits_{eE/\Lambda^2}^\infty dy
{\exp \left( \frac{\sigma^2}{eE}(\varphi-y) \right)
\over {(1+y^2)^2(y+\pi n -\varphi)^3}}
$$
$$
\times \biggl[ y(y^2+4)(\pi n -\varphi +y)
{\sigma^2\over eE}+ 6(\pi n-\varphi)-2y(y^2-5)\biggl].
\eqno(49)
$$

However this cut-off regularization with a finite $\Lambda$
 is not the only one that can be performed.
Instead of it, it is also possible to send $\Lambda \to \infty$
and simultaneously rewrite our formula for the EP
in $4- 2\epsilon$
dimension, introducing therefore some kind of dynamical regularization
which seems to be more useful for numerical simulations.
Then, for weak EF the EP can be expressed
in the following, more convenient form:
$$
V^{(4D)}_{eff}(\sigma)={\sigma^2 \over 2 \lambda} -
\frac{(eE)^2}{8\pi^2}\Gamma(-1 + \epsilon)
\left[4\zeta(-1 + \epsilon, -i{\sigma^2\over 2eE}) +
 i{\sigma^2 \over eE}\right]
$$
$$
-{R \sigma^2 \over 96\pi^2}\Gamma(-1+\epsilon)+
{iR(eE)^{2/3} \over 48\pi^3 3^{1/3}}
\exp(-\pi{\sigma^2 \over eE})\Gamma(\frac{2}{3})\sigma^{2/3}.
\eqno(50)
$$

Figure 4 presents the real part of EP  in the
$(4-2\epsilon)$ -dimensional  case in flat spacetime. It has been found by the
numerical analysis that the final results are almost independent of 
the special $\epsilon$ values. 
The phase transition is of second-order, because there exists 
a unique critical EF strength, defined as the value for which 
the zero extremum becomes 
 a maximum and, simultaneuosly, a nonzero absolute minimum appears.
This critical value separates the situations where the zero extremum is
a maximum and a non-zero absolute minimum of the potential exists, from
the ones where the origin is the only and absolute minimum.

Figures 5 and 6 show the shapes of the real part of EP in curved
$(4-2\epsilon)$-dimensional spacetime for fixed EF 
strength and spacetime
curvature on the abscissa axis. In both cases, a second-order phase transition
occurs. 

It should be emphasized that in the presence of an imaginary part in the
EP, the criterion for restoration of
the chiral symmetry  must be modified
\cite{Jp, K1, JpEF}.
In fact, in that case the order parameter has to be
chosen as $\sigma \langle {\overline \psi}\psi \rangle$, but not 
simply as
$\sigma$ because, although the later vanishes at the critical point,
the former might still break the chiral symmetry through the 
corresponding term in the action (7).
Formula (11) gives that
$$
\langle {\overline \psi}\psi \rangle = iN\Im V'_{eff}(\sigma) -
\frac{N}{\lambda}\sigma
\eqno(51)
$$
and, therefore, we should only check whether our EP obeys
the criterion:
$$
\lim_{\sigma \rightarrow 0} \sigma \Im V'_{eff}(\sigma) = 0.
\eqno(52)
$$ 
Formulae (43) and (50) make us sure that this criterion is
satisfied here for both the 3D and 4D cases.

\section{Conclusions.}

Phenomenological models with four-fermion interaction seem to be
useful for the description of low- and inermediate-energy physics of strong
interactions \footnote{More detailed information and complete literature
list can be found in \cite{PhysRep}, for example}. The fermions are
usually treated as quarks and the composite particles generated by the
dynamical symmetry breaking as mesons, despite of some simplifications
making the models under consideration more calculable.

The nonzero vacuum expectation value of composite bifermionic
operator $\langle \overline{\psi} \psi \rangle$ plays in DSB scheme
just the same role as $\langle \varphi \rangle$ in the traditional Higgs one
and defines the dynamically generated mass of fermions in our case.  

Usually the symmetry to be broken
under the DSB mechanism is the chiral one. A dynamical version of fermions
mass generation and dynamical chiral symmetry breaking have been 
investigated very carefully and some fruitful applications
for the real high-energy physics \cite{RWP}-\cite{GMS} have been found. 

In particular, the NJL model in external electromagnetical field has been
studied
from  different points of view to realize the crucial role of that field
in the DSB for flat spacetime \cite{Sch}- \cite{naf}. The effect of
dynamical chiral symmetry breaking by external magnetic field at any
attractive interaction of the fermions has been
shown to be universal and model independent phenomenon \cite{GMS}.

In our paper we have extended these investigations to the case of the
presence of an electric field  and non-zero spacetime curvature. Both 
these values try to restore the chiral symmetry in contrast to the magnetic   
field and their competition with the DSB catalysis by the magnetic field
has to be understood more precisely.

The Gross-Neveu model with the simplest four-fermionic interaction
$(\overline{\psi} \psi)^2$ is invariant under the discrete chiral   
transformation
$$
\psi \to \gamma_5 \psi
\eqno(54)
$$
 Because  of the discrete type of
this symmetry Goldstone modes do not  appear.

However NJL model has a continuous chiral invariance under
the transformations
$$
\psi \to e^{i\theta\gamma_5}\psi
\eqno(55)
$$

In contrast
to the previous case Goldstone bosons have to appear if only the
symmetry is being broken.
In the phenomenological applications of NJL model $\pi$ field is treated
as Goldstone modes corresponding to the massless pions,  which obtain the
kinetic terms through the
quantum corrections. Of course, if we take into account the quarks' current
masses these modes become pseudo-Goldstone ones \cite{RWP, PhysRep}.
We only split them out of our consideration from
the very beginning by the appropriate choice of the direction in the
$\sigma-\pi$ space and use this choice's advantage that DSB is ruled by
$\sigma$ itself.

The investigation of the phenomenon realization in the presence of
magnetic field in flat spacetime has been performed both for NJL model and
for quantum electrodynamics and quantum excitations of $\sigma$ and $\pi$
fields  dispersion laws have been
found \cite{GMS}.  To generalize this calculations for the curved
spacetime case is an interesting problem we are dealing with now.

However, here we have studied the phase structure of the NJL model
in curved spacetime with
external constant pure electric field only. In three dimensions, a
first-order phase transition takes place, both on the electric
 field strength and
on the spacetime curvature. In the four-dimensional case -- or, more
precisely, in $4-2\epsilon$ dimensions with negligible $\epsilon$ -- the
 phase transition is a typical  second-order one, on both of these
external parameters.

Our approximation is quite consistent, because all of the critical values
are very small in comparison with the characteristic scale of the model $\mu$.
It should be noted that the consideration, in a nonperturbative way, 
  of an external electric field 
provides very interesting effects even in flat spacetime, and that the linear
curvature terms induce  some additional processes of chiral symmetry
restoration. 

We clearly observe that a positive spacetime curvature tries
 to restore chiral symmetry,
as an external electric field does. This is in contrast with what happens in
the case of a magnetic filed, where
chiral symmetry is broken for any value of the field strength although
positive curvature might restore chiral symmetry again for some critical
value.

\section {Acknowledgments} 
 
We thank A.A. Andrianov and S. D. Odintsov for useful discussions.
This work has been partly financed by DGICYT (Spain), project PB93-0035, and
by  CIRIT (Generalitat de Catalunya),  grant 1995SGR-00602.
The work of Yu.I.Sh.  was supported in part by Ministerio de
Educaci\'on y Cultura (Spain), grant SB96-AN4620572. Yu.I.Sh. also
expresses
 his deep gratitude to A. Letwin and R. Patov for their kind support.

\pagebreak
\begin{figure}[tb]
 \centerline{\epsfxsize=12cm \epsfbox{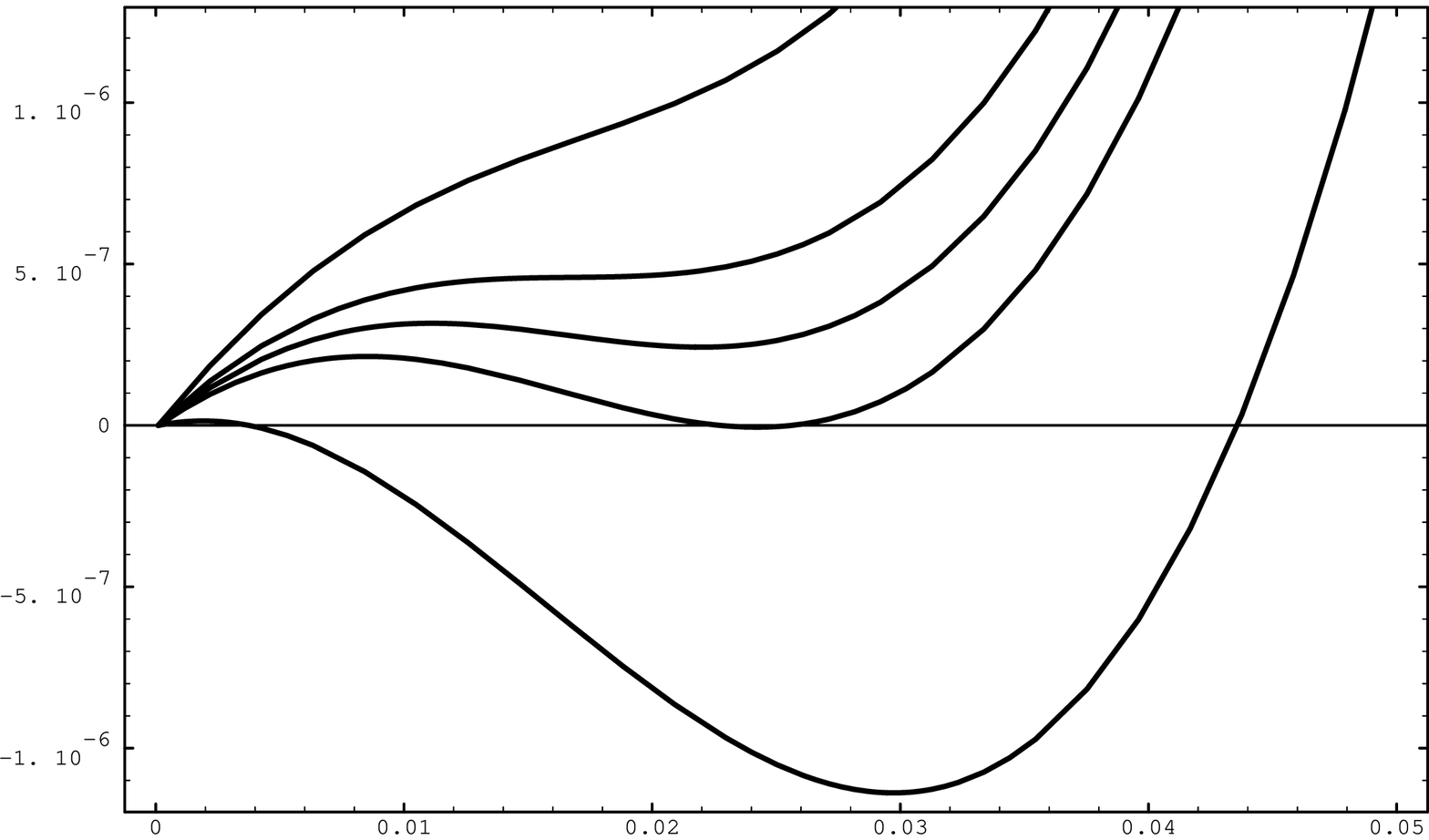}}
\vskip-28mm
\caption{{\protect\small
The behaviour in 3D of the real part of the effective potential
 $\Re V_{eff, R}^{flat}/\mu^3$ as a function
of $\sigma/\mu$ is depicted for
$R=0$ and fixed $\lambda\mu= -100$.   
The curves from the upper part to the bottom of the plot correspond to
the following electric   field strengths:
$eE/\mu^2=0.0003; 0.0023; 0.00020; 0.00017; 0.00005$, respectively.
The critical values, defined as usual, are given by:
$eE_{c1}/\mu^2=0.00023$; $eE_{c}/\mu^2=0.00017$;
$eE_{c2}/\mu^2=0.00005$.}}
\end{figure}
\pagebreak
\begin{figure}[bt]
 \centerline{\epsfxsize=12cm \epsfbox{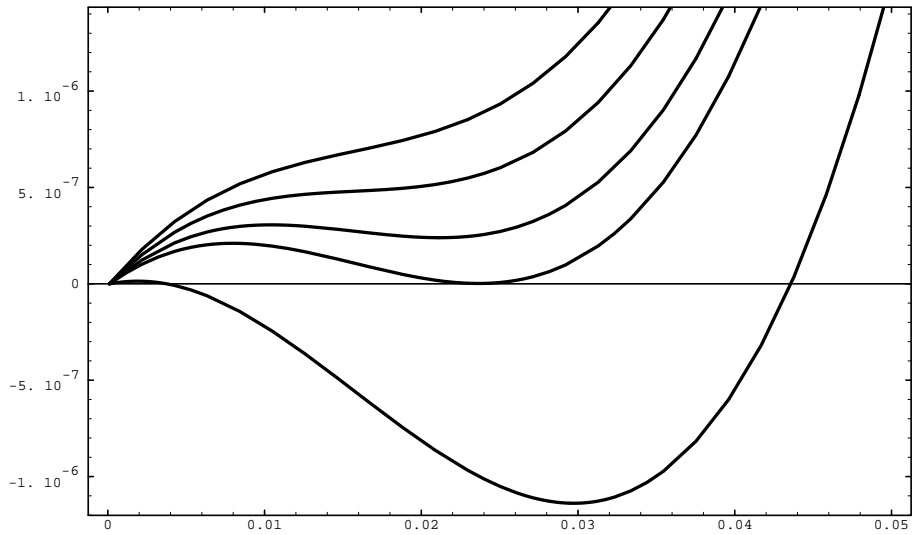}}
\vskip-28mm
\caption{{\protect\small
Behaviour in 3D of $\Re V_{eff, R}/\mu^3$
as a function
of $\sigma/\mu$  for
fixed $eE/\mu^2= 0.00005$ and  $ \lambda\mu= -100$.
From above to below,  the curves in the  plot correspond to
the following values of
$R/\mu^2=0.006; 0.005; 0.004; 0.0032; 0$, respectively.
The critical values, defined as usual, are given by:
$R_{c1}/\mu^2=0.005$; $R_{c}/\mu^2=0.0032$;
 $R_{c2}/\mu^2=0$.}}
\end{figure}
\pagebreak
\begin{figure}[htb]
 \centerline{\epsfxsize=12cm \epsfbox{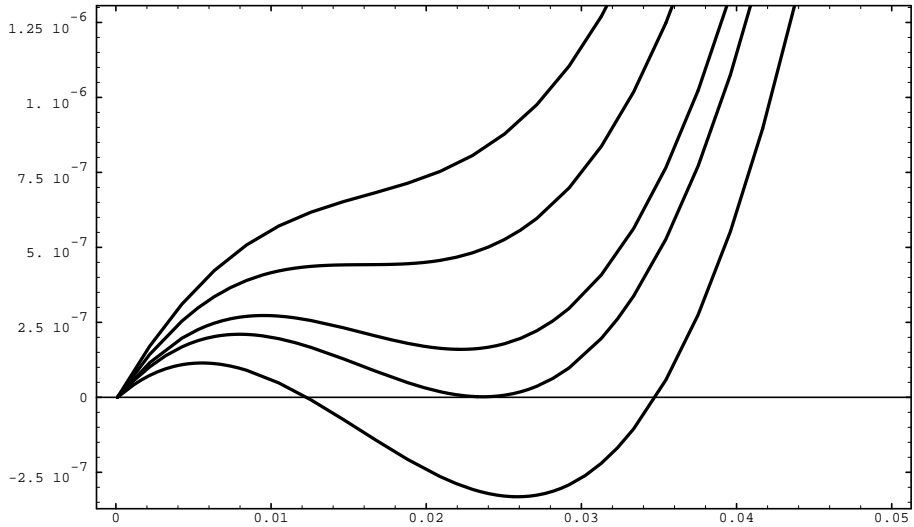}} 
\vskip-28mm
\caption{{\protect\small
Behaviour in 3D of $\Re V_{eff, R}/\mu^3$
as a function
of $\sigma/\mu$, for
fixed $R/\mu^2=0.0032$ and $\lambda\mu= -100$.
Starting from above, the curves correspond to 
the following values of
$eE/\mu^2=0.00015; 0.00011; 0.00007; 0.00005; 0.00001$, respectively.
The critical values, defined as usual, are given by:
$eE_{c1}/\mu^2=0.00011$; $eE_{c}/\mu^2=0.00005$,
while $eE_{c2}/\mu^2$ does not exist for the given values of
 $\mu\lambda$ and  $R/\mu^2$.} }
\end{figure}
\pagebreak
\begin{figure}[htb]
 \centerline{\epsfxsize=12cm \epsfbox{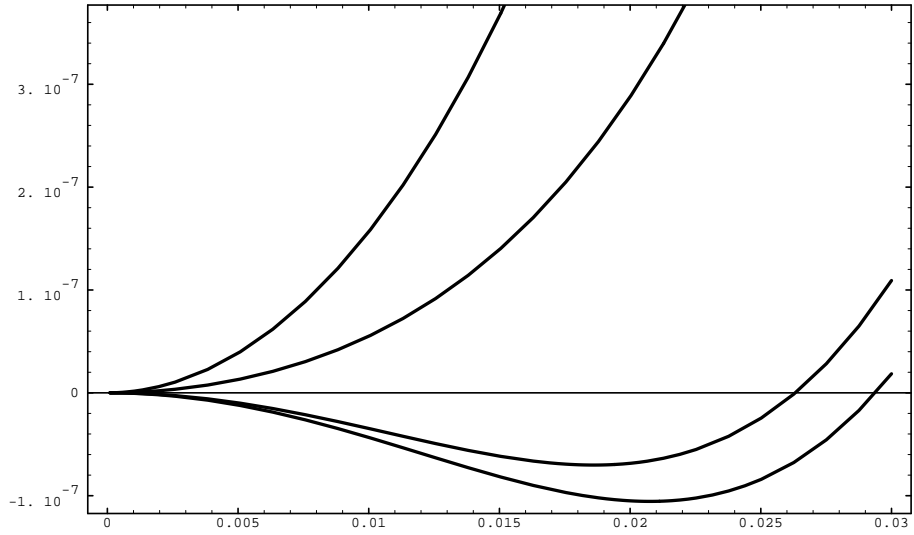}}
\vskip-28mm
\caption{{\protect\small
Behaviour of
$\Re V_{eff}/\mu^{(4-2\epsilon)}$
 in a $(4-2\epsilon)$-dimensional spacetime as a function
of $\sigma/\mu$, for
fixed $R = 0$, $\epsilon = 0.02$ and $\lambda\mu= -1000$.
Starting from above, the curves in the plot correspond to
the following values of
$eE/\mu^2=0.02; 0.01; 0.005; 0.001$, respectively.
The critical value is given by:
$eE_{c}/\mu^2=0.02$.} }
\end{figure}
\pagebreak
\begin{figure}[htb]
 \centerline{\epsfxsize=12cm \epsfbox{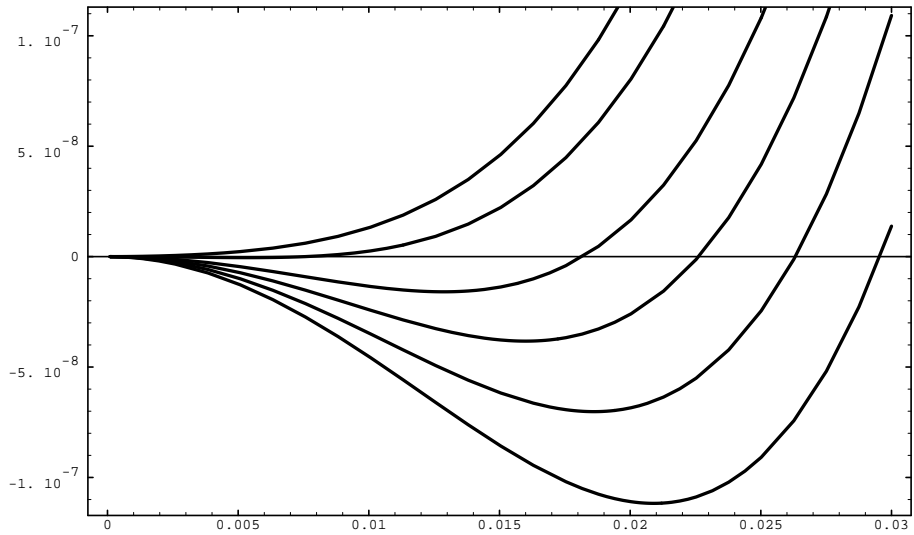}}
\vskip-28mm
\caption{{\protect\small
Behaviour of
$\Re V_{eff}/\mu^{(4-2\epsilon)}$
 in a $(4-2\epsilon)$-dimensional spacetime
as a function
of $\sigma/\mu$, for
$eE/\mu^2= 0.001$, $\epsilon = 0.005$
 and fixed $\lambda\mu= -1000$.
Starting from above, the curves correspond to
the following values of the curvature:
$R/\mu^2 = 0.0045; 0.0035; 0.02; 0.01; 0; -0.001$, respectively.
The critical value is obtained for
$R{c}/\mu^2=0.0035$.}}
\end{figure}
\pagebreak
\begin{figure}[htb]
 \centerline{\epsfxsize=12cm \epsfbox{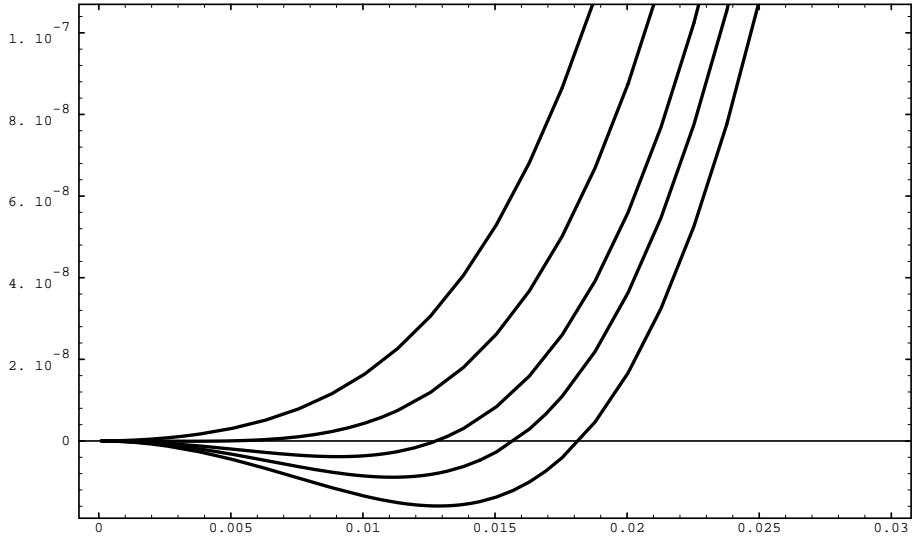}}
\vskip-28mm
\caption{{\protect\small
Behaviour of
$\Re V_{eff}/\mu^{(4-2\epsilon)}$
 in a $(4-2\epsilon)$-dimensional spacetime,
as a function
of $\sigma/mu$, depicted for
$R/\mu^2= 0.002$, $\epsilon = 0.005$
 and fixed $\lambda\mu= -1000$.
From above, the curves in the plot correspond to
the following values of the electric field strength:
$eE/\mu^2 = 0.004; 0.0028; 0.02; 0.015; 0.001$, respectively.
The critical value is reached at
$eE_{c}/\mu^2=0.0028$.}}
\end{figure}
\end{document}